AI Model for Predicting Binding Affinity of Antidiabetic Compounds Targeting PPAR

La Ode Aman[1]*, Aiyi Asnawi[2]

[1] Department of Pharmaceutical Chemistry, Faculty of Sports and Health, Universitas Negeri Gorontalo, Jl. Jenderal Sudirman No 6 Gorontalo 96128, Indonesia
[2] Department of Pharmacochemistry, Faculty of Pharmacy, Universitas Bhakti Kencana, Jl. Soekarno-Hatta No. 754, Bandung 40617, Indonesia

Corresponding author: laode_aman@ung.ac.id

**Abstract**

This study aims to develop a deep learning model for predicting the binding affinity of ligands targeting the Peroxisome Proliferator-Activated Receptor (PPAR) family, using 2D molecular descriptors. A dataset of 3,764 small molecules with known binding affinities, sourced from the ChEMBL database, was preprocessed by eliminating duplicates and incomplete data. Molecular docking simulations using AutoDock Vina were performed to predict binding affinities for the PPAR receptor family. 2D molecular descriptors were computed from the SMILES notation of each ligand, capturing essential structural and physicochemical features. These descriptors, along with the predicted binding affinities, were used to train a deep learning model to predict binding affinity as a regression task. The model was evaluated using metrics such as Mean Squared Error (MSE), Mean Absolute Error (MAE), and R-squared ($R^2$). Results indicated strong performance with an $R^2$ value of 0.861 for the training set and 0.655 for the test set, suggesting good model generalization. The model shows promise for predicting ligand-receptor interactions and can be applied in drug discovery efforts targeting PPAR-related diseases.

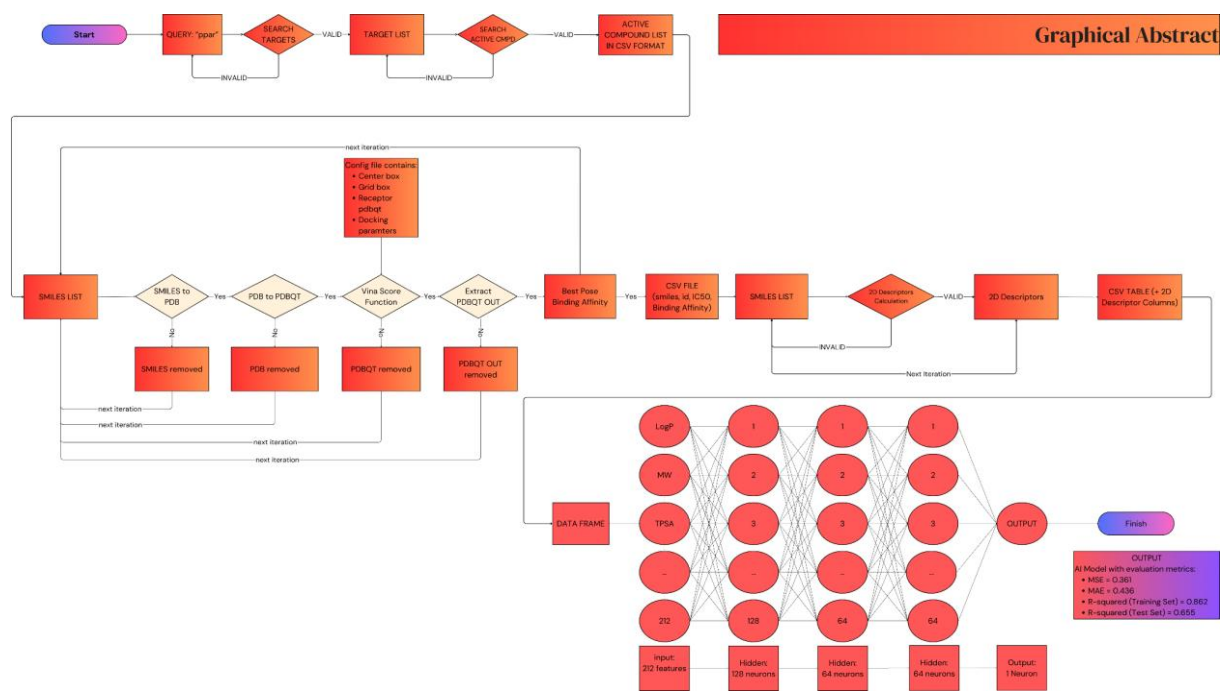

**Introduction**

Diabetes mellitus has emerged as a significant global health concern, with its prevalence escalating alarmingly across both developed and developing nations. The International Diabetes Federation reported that approximately 537 million adults were living with diabetes in 2021, a figure projected to rise substantially in the coming years [1]. This chronic condition is characterized by elevated blood glucose levels due to insulin resistance and inadequate insulin production, leading to severe complications such as cardiovascular diseases, renal failure, and increased mortality rates [2][3]. The economic burden of diabetes is also profound, with healthcare costs associated with the disease reaching an estimated $245 billion in the United States alone in 2012 [4][5]. The increasing incidence of diabetes necessitates the development of effective antidiabetic therapies to mitigate its health and economic impacts.

Peroxisome proliferator-activated receptors (PPARs) play a crucial role in glucose metabolism and insulin sensitivity, making them a promising target for therapeutic interventions in type 2 diabetes (T2D). PPARs, which include three isoforms—PPAR-α, PPAR-β/δ, and PPAR-γ—are nuclear receptors that regulate gene expression involved in lipid and glucose metabolism [6]. Agonists of PPAR-γ, such as thiazolidinediones, have been clinically utilized to improve insulin sensitivity and manage hyperglycemia associated with T2D [7]. Additionally, PPAR-δ has been implicated in enhancing glucose homeostasis through pathways involving adiponectin and sodium-glucose transporters, highlighting the multifaceted roles of PPARs in metabolic regulation [8][9]. The therapeutic potential of PPARs extends beyond glucose metabolism, as they also influence inflammatory processes and energy balance, which are critical in the pathophysiology of diabetes [10][11].

Despite the promising role of PPARs in diabetes management, the drug discovery process for antidiabetic compounds faces significant challenges. The complexity of PPAR-ligand interactions presents a substantial hurdle, as the efficacy of PPAR agonists can be influenced by their selective activation of different receptor isoforms, leading to varied metabolic outcomes [12][6]. Furthermore, the costs associated with experimental methods for drug development are considerable, often limiting the exploration of novel therapeutic agents [4][13]. The intricate nature of diabetes pathophysiology, combined with the economic constraints of drug development, underscores the need for innovative strategies to enhance the discovery and efficacy of antidiabetic therapies.

The integration of artificial intelligence (AI) in predicting binding affinity represents a transformative shift in drug discovery, offering a cost-effective alternative to traditional experimental methods. Traditional approaches to determining binding affinity often involve extensive laboratory work, which can be time-consuming and resource-intensive. In contrast, AI models leverage vast datasets and computational power to predict how well a drug candidate will bind to its target, significantly reducing the time and costs associated with drug development Singh [14][15]. By employing machine learning algorithms, researchers can analyze complex interactions between proteins and ligands, leading to more efficient identification of promising drug candidates [16][17].

Success stories abound in the application of AI models for molecular docking, dynamics, and quantitative structure-activity relationship (QSAR) analyses. For instance, recent advancements in deep learning have enabled the development of scoring functions that improve the accuracy of binding affinity predictions in protein-ligand interactions [18][19]. These models have demonstrated their potential in various studies, such as the identification of the flavonoid Troxerutin as a candidate TRPV1 antagonist, showcasing the capability of AI to streamline the drug discovery process [20]. Furthermore, AI-driven molecular generation techniques have been employed to create novel compounds with desirable properties, enhancing the efficiency of lead optimization [21][22]. The ability of AI to predict binding affinities accurately has been validated through numerous studies, underscoring its role in optimizing lead compounds and advancing drug discovery efforts [23][24].

Accurate binding affinity prediction is paramount for optimizing lead compounds in drug discovery. The binding affinity of a ligand to its target protein is a critical determinant of its efficacy and safety profile. Models that can reliably predict these affinities enable researchers to prioritize compounds for further development, thus enhancing the likelihood of clinical success [25][26]. Moreover, the ability to predict how modifications to a compound affect its binding affinity allows for iterative design processes that can lead to the identification of more potent and selective drug candidates [27][28]. As AI continues to evolve, its integration into drug discovery processes will likely lead to more rapid advancements in therapeutic development, ultimately improving patient outcomes.

The application of artificial intelligence (AI) in predicting binding affinity for peroxisome proliferator-activated receptors (PPAR)-ligand systems presents several limitations that hinder the effectiveness of current models. One significant challenge is the limited size of datasets available for training these models. The performance of AI algorithms heavily relies on the quantity and quality of data; however, the datasets specific to PPAR-ligand interactions are often small and may not encompass the diversity of chemical structures and biological contexts necessary for robust model training Libouban [29][30]. This scarcity can lead to overfitting, where models perform well on training data but fail to generalize to unseen compounds, thereby limiting their predictive power [31].

Another critical limitation is the feature selection process used in these AI models. Many existing models primarily focus on a narrow range of features, such as molecular descriptors or 3D structural information, which may not capture the complexity of PPAR-ligand interactions adequately [32]. For instance, while some models utilize advanced techniques like deep learning to analyze structural data, they often overlook essential biochemical properties that influence binding affinity, such as solubility, permeability, and metabolic stability [32]. This lack of comprehensive feature representation can result in inaccurate predictions and hinder the optimization of lead compounds targeting PPARs.

Moreover, the generalizability of AI models is a significant concern, particularly for antidiabetic compounds targeting PPARs. Many current models have been developed with a focus on specific protein-ligand pairs or classes of compounds, which may not translate effectively to other PPAR-ligand systems [31]. The absence of comprehensive models specifically designed for antidiabetic compounds targeting PPARs further exacerbates this issue. Most existing methodologies do not account for the unique pharmacological profiles and mechanisms of action associated with PPAR agonists and antagonists, limiting their applicability in drug discovery for diabetes [33]. This gap underscores the need for dedicated AI frameworks that can integrate diverse datasets and accurately model the interactions specific to PPAR-ligand systems.

The primary objective of this research is to develop and validate an AI-based model for accurately predicting the binding affinity of antidiabetic compounds targeting PPARs (Peroxisome Proliferator-Activated Receptors). The model aims to leverage advanced molecular descriptors, cutting-edge machine learning architectures, and robust datasets to improve prediction accuracy and generalizability. By integrating computational chemistry insights with AI techniques, this research seeks to provide a reliable tool for screening and optimizing potential drug candidates.

This research holds significant potential to revolutionize antidiabetic drug discovery by streamlining and enhancing the development process. By providing a robust computational framework, it accelerates the identification of promising lead compounds, thereby reducing the time required to advance potential therapeutics through the early stages of drug development. Moreover, the reliance on in silico predictions minimizes the need for costly experimental methods, such as labor-intensive in vitro and in vivo assays, making the process more economically feasible.

Beyond cost and efficiency, this research also advances the integration of AI technologies into precision medicine, enabling the design of tailored therapeutic approaches. This targeted strategy enhances the management of diabetes, addressing the unique needs of individual patients more effectively. By overcoming the limitations of traditional drug discovery methodologies, this work paves the way for innovative and affordable solutions, aligning with the broader vision of leveraging AI to transform healthcare and improve treatment outcomes for diabetes patients.

**Methods**

**Computational Infrastructure**

To ensure efficient and accurate execution of this study, a combination of high-performance hardware and state-of-the-art software tools was utilized for data preparation, molecular docking simulations, and deep learning model development.

**Hardware Configuration**

The majority of computational tasks, excluding molecular docking simulations, were executed on a Hewlett-Packard HP Z840 workstation. This system, equipped with a 40-core Intel® Xeon® E5-2650 v3 processor, 32.0 GiB of RAM, and an NVIDIA GeForce RTX™ 3070 GPU, provided the processing power required for demanding tasks such as descriptor generation and neural network training. Its 3.0 TB storage capacity ensured efficient data handling and storage throughout the project. These specifications were instrumental in managing large datasets and performing computationally intensive operations.

Molecular docking simulations, which required exceptional computational precision, were conducted on the Fugaku Supercomputer. This cutting-edge facility offered unparalleled performance, ensuring highly accurate ligand-receptor binding affinity calculations and enabling the study to achieve precise docking results.

*Software Environment*

The computational setup for this study was based on the Ubuntu 24.10 (64-bit) operating system, supported by the Linux kernel version 6.11.0-9-generic. The GNOME 47 desktop environment with the Wayland windowing system provided a stable and efficient user interface. This configuration was chosen to ensure compatibility and optimal performance with the required scientific software tools.

*Key Computational Tools*

Python 3.11 served as the central programming platform, supplemented by specialized libraries to streamline specific tasks. RDKit (Version 2024.09.1) was used for cheminformatics operations, including molecular fingerprint generation and SMILES processing. Open Babel (Version 3.1.1) facilitated molecular file conversions to ensure interoperability. TensorFlow (Version 2.13) was employed for developing and training deep learning models, while Scikit-learn (Version 1.4.1) supported preprocessing and performance evaluation. Pandas (Version 2.2.0) enabled efficient data manipulation.

Molecular docking simulations were carried out using AutoDock Vina (Version 1.2.5) on the Fugaku Supercomputer, which allowed precise prediction of binding affinities. Data visualization was performed with Matplotlib (Version 3.8.1) and Seaborn (Version 0.13.0), which provided detailed and aesthetically pleasing graphical representations of results.

**Macromolecule and Inhibitor Identification**

Protein targets were retrieved from the ChEMBL database using the keyword "PPAR," corresponding to the Peroxisome Proliferator-Activated Receptor family. This search identified key isoforms, including PPARα, PPARγ, and PPARδ, which play crucial roles in regulating lipid and glucose metabolism, making them significant targets for antidiabetic therapies.

Small molecule inhibitors were subsequently extracted by filtering for compounds with IC50 values, ensuring that the dataset included bioactivity data necessary for accurate modeling. Compounds lacking IC50 data or showing insufficient potency (IC50 above a defined threshold) were excluded to focus on the most promising inhibitors. The Python scripts used to perform these queries and extractions are provided in Appendices 1 and 2.

Data Preparation and Cleaning

A meticulous data-cleaning process was employed to enhance the quality and reliability of the dataset. Duplicate entries, identified by identical molecule_chembl_id values, were consolidated by retaining the record with the lowest IC50 value, which indicates higher potency. Rows containing missing or invalid IC50 values, such as NaN or zero, were removed to prevent inaccuracies in downstream analyses. Additionally, columns irrelevant to the computational workflows, such as pref_name and search_term, were eliminated to streamline the dataset.

These steps produced a high-quality dataset, optimized for molecular descriptor generation, docking simulations, and machine learning model development. Detailed descriptions of the data-cleaning procedures are included in Appendix 3.

This systematic approach, combining advanced data processing techniques and robust computational tools, ensured the reliability and precision of the study's outcomes.

Binding Affinity Prediction Using AutoDock Vina

Binding affinities were predicted using AutoDock Vina, a widely adopted molecular docking tool designed to model the interactions between ligands and receptors. This involved careful preparation of both the macromolecule (receptor) and the small molecules (ligands) to ensure accurate and biologically relevant simulations.

**Receptor Preparation**

The 3D structure of the PPARγ protein (PDB ID: 7VWG) was used as the receptor in this study. Prior to docking, the protein structure was refined to ensure its stability and functional integrity. Water molecules, ions, and other non-essential elements were removed to prevent interference with ligand binding. The binding site was identified and defined using a grid box centered at coordinates X = -23.234, Y = -18.882, Z = 9.755, with dimensions X = 20, Y = 20, and Z = 20. This configuration ensured that the docking simulations were focused on the receptor's active site, capturing biologically meaningful interactions.

To validate the binding site, a re-docking procedure was conducted with the co-crystallized ligand from PDB ID 7VWG. This step verified the accuracy of the docking protocol, ensuring reliable predictions for subsequent simulations.

**Docking Simulation Setup**

AutoDock Vina was configured to achieve high accuracy in predicting ligand-receptor interactions. The grid spacing was set to 0.375 to provide a fine resolution for the search space. The exhaustiveness parameter was adjusted to 32 to thoroughly explore potential binding conformations, balancing precision and computational efficiency. To optimize performance, the simulations were executed using 32 CPU threads.

The Vina scoring function was employed to estimate binding affinities, ranking ligands based on their predicted interaction strength with the PPARγ receptor. This systematic approach ensured reliable predictions, forming the basis for evaluating the potential of candidate antidiabetic compounds.

This comprehensive docking procedure, combining precise receptor preparation, grid definition, and advanced computational settings, provided a robust framework for estimating ligand binding affinities in the study.

**Docking Results and Pose Validation**

The most favorable docking pose targeting the PPAR receptor yielded a binding affinity of -10.09 kcal/mol. The results demonstrated a high level of agreement between the docked ligand pose and the native conformation, affirming the reliability of the docking procedure. Figure 1 illustrates the superimposition of the docked ligand (orange carbon) and its native conformation (blue carbon) within the PPAR binding site, highlighting their overlap and alignment.

*Figure 1. Superimposition of the docked native ligand (orange carbon) and its original conformation (blue carbon) within the PPAR receptor's active site.*

**Molecular Descriptor Calculation**

To numerically represent the chemical and structural features of the ligands, 2D molecular descriptors were calculated using the RDKit library. These descriptors encompassed topological, geometric, and electronic properties, enabling comprehensive molecular characterization.

Selected descriptors were tailored to capture the relevant molecular attributes associated with ligand-receptor interactions. Each molecule's descriptor values were formatted as feature vectors, forming the input dataset for subsequent model training. Details of descriptor generation and the corresponding Python code are outlined in Appendix 5.

**Deep Learning Model Construction for PPAR Binding Affinity Prediction**

A deep learning model was developed to predict the binding affinities of small molecules targeting PPAR. Using 2D molecular descriptors as input features, the model provided a robust framework for capturing the complex relationships between molecular properties and binding strength.

Data Preparation

The descriptors served as input features (X), while binding affinities (y) were used as the target variable. For training and validation, the dataset was split into 80% for training and 20% for testing. Binding affinity values were scaled to enhance numerical stability during training.

Model Architecture

The predictive model was developed as a feedforward neural network designed to accommodate the descriptor feature vectors as input. The network architecture consisted of three hidden layers, each responsible for capturing the complex, non-linear interactions between the molecular descriptors. The first hidden layer contained 128 neurons, followed by a second and third hidden layer with 64 neurons each. To introduce non-linearity and enhance the model's ability to learn intricate patterns, the ReLU activation function was applied to all hidden layers. The network's output layer consisted of a single neuron, which provided the predicted binding affinity values, making the model suitable for regression tasks. This architecture was carefully chosen to ensure that the model could effectively predict the binding affinities of the ligands targeting the Peroxisome Proliferator-Activated Receptor (PPAR) family based on their molecular descriptors.

Training and Evaluation

The model was trained using the Adam optimizer with the mean squared error (MSE) loss function over five epochs and a batch size of 32. Validation data comprising 20% of the training set was used to monitor performance during training.

Model evaluation involved calculating metrics such as Mean Squared Error (MSE), Mean Absolute Error (MAE), and R-squared ($R^2$) values. These metrics quantified prediction accuracy, error magnitude, and variance explained by the model for both training and testing subsets. A complete implementation of the workflow, including Python scripts, is provided in Appendix 6.

**Results**

The search for macromolecular targets using the term "PPAR" identified proteins associated with the Peroxisome Proliferator-Activated Receptor (PPAR) family. Each target was categorized by its respective CHEMBL ID (target_chembl_id) and the associated preferred name (pref_name), representing individual receptors within the PPAR family.

**Table 1.** Macromolecular targets associated with the PPAR family identified through the "PPAR" keyword search.

| target_chembl_id | pref_name |
|---|---|
| CHEMBL3559683 | Peroxisome proliferator-activated receptor |
| CHEMBL239 | Peroxisome proliferator-activated receptor alpha |
| CHEMBL3979 | Peroxisome proliferator-activated receptor delta |
| CHEMBL235 | Peroxisome proliferator-activated receptor gamma |
| CHEMBL6116 | Peroxisome proliferator-activated receptor gamma coactivator 1-alpha |
| CHEMBL2095162 | Peroxisome proliferator-activated receptor gamma/Nuclear receptor coactivator 1 |
| CHEMBL2095163 | Peroxisome proliferator-activated receptor gamma/Nuclear receptor coactivator 2 |
| CHEMBL2095161 | Peroxisome proliferator-activated receptor gamma/Nuclear receptor coactivator 3 |
| CHEMBL2096976 | Peroxisome proliferator-activated receptor gamma/Nuclear receptor corepressor 2 |

Identifying Small Molecules as PPAR Modulators

The subsequent phase focused on identifying small molecule modulators targeting the PPAR receptor family. This involved querying the ChEMBL database using the target_chembl_id identifiers retrieved from the macromolecule search. The dataset was curated to include only compounds with available IC50 values, ensuring the inclusion of bioactivity data relevant for evaluating modulator potency.

A total of 15,161 small molecules were identified as potential modulators of the PPAR receptor family. The IC50 values, recorded in the standard_value column, served as the key metric for determining compound efficacy, with lower IC50 values signifying stronger modulatory activity. These values are expressed in nanomolar (nM) units.

A representative subset of the dataset is provided in **Table 2**, showcasing select modulators along with their IC50 values and supplementary details. The pref_name column highlights the specific PPAR receptor targeted by each compound, with a particular focus on Peroxisome Proliferator-Activated Receptor Gamma (PPAR-γ). Additionally, the canonical_smiles column contains the SMILES notation of each compound's molecular structure, enabling further computational analyses such as molecular docking and the extraction of features for machine learning-based predictions.

**Table 2.** A sample of small molecule modulators targeting the PPAR receptor family, including IC50 values and SMILES representations.

| molecule_chembl_id | canonical_smiles | standard_value | pref_name |
|---|---|---|---|
| CHEMBL327767 | CCCc1cc(Oc2ccc(CC(C)C)cc2)ccc1OCCCOc1cccc(C2SC(=O)NC2=O)c1 | 10000 | Peroxisome proliferator-activated receptor alpha |
| CHEMBL94496 | CCCc1cc(Oc2ccc(C(C)C)cc2)ccc1OCCCOc1cccc(C2SC(=O)NC2=O)c1 | 2100 | Peroxisome proliferator-activated receptor alpha |
| CHEMBL420441 | CCCc1cc(Oc2ccc(Cl)cc2)ccc1OCCCOc1cccc(C2SC(=O)NC2=O)c1 | 100 | Peroxisome proliferator-activated receptor alpha |
| CHEMBL121 | CN(CCOc1ccc(CC2SC(=O)NC2=O)cc1)c1ccccn1 | 50000 | Peroxisome proliferator-activated receptor alpha |
| CHEMBL330191 | CCCc1cc(Oc2ccccc2)ccc1OCCCCOc1cccc(C2SC(=O)NC2=O)c1 | 5000 | Peroxisome proliferator-activated receptor alpha |
| CHEMBL300629 | CCCc1cc(Oc2ccccc2)ccc1OCCCOc1ccc(C2SC(=O)NC2=O)cc1 | 50000 | Peroxisome proliferator-activated receptor alpha |

| CHEMBL328615 | CCCc1cc(Oc2ccc(Cl)c(C)c2)ccc1OCCOc1cccc(C2SC(=O)NC2=O)c1 | 162 | Peroxisome proliferator-activated receptor alpha |

The dataset, comprising 3,764 small molecules with associated IC50 values, was prepared for subsequent steps, including data preprocessing, molecular docking simulations, and the construction of machine learning models for binding affinity prediction. Detailed Python implementation for this process is provided in Appendix 2.

Data Preprocessing

The dataset from Step 2, initially containing 3,764 small molecules, underwent a comprehensive cleaning process to ensure its suitability for further analysis. Duplicate entries with identical molecule_chembl_id were identified and removed. From the duplicates, only the entries with the lowest standard_value, indicating the most potent inhibitors, were retained to preserve the relevance and accuracy of the dataset.

Additionally, entries with missing or invalid values in the standard_value column, such as NaN or 0, were discarded. These incomplete or erroneous data points could not contribute meaningful insights for binding affinity predictions, and their exclusion improved the overall quality of the dataset.

Columns not required for molecular docking or machine learning tasks, such as pref_name and search_term, were also removed to streamline the dataset for the computational workflows.

After these cleaning steps, the dataset was refined to 2,191 unique small molecules, formatted and prepared for molecular docking and the creation of predictive deep learning models. The Python code used for this data cleaning process is provided in Appendix 3.

Binding Affinity Prediction Using AutoDock Vina

The binding affinities of the 3,764 small molecules, cleaned in Step 3, were predicted using molecular docking with AutoDock Vina. The process began with preparing each ligand from the cleaned dataset, using its canonical SMILES string. These SMILES strings were converted into the PDBQT format required for docking simulations in AutoDock Vina. This conversion was performed using RDKit and Open Babel, with the corresponding Python code provided in Appendix 4.

Following ligand preparation, molecular docking simulations were carried out using AutoDock Vina. Each ligand was docked with a receptor protein representing the PPAR target. The output from these simulations provided binding affinity predictions, which were extracted from the "REMARK VINA RESULT" section of the output file, indicating the optimal docking score.

The predicted binding affinities, measured in kcal/mol, reflect the strength of the interaction between each ligand and its respective receptor. These results were compiled into a new dataset and saved in CSV format for subsequent analysis. The dataset includes key columns such as molecule_chembl_id (unique identifier for each ligand), canonical_smiles (SMILES string for the ligand's structure), standard_value (IC50 value in nM), and binding_affinity (the predicted binding affinity in kcal/mol from the docking simulations).

An example of the docking results is shown in Table 3, where several small molecules are listed along with their IC50 values and predicted binding affinities. The receptor used in the docking simulations was a PPAR protein model in PDBQT format, and the final docking results were stored in the output file *ppar_docking.csv*.

**Table 3.** Example of docking results, showing small molecules, IC50 values, and predicted binding affinities.

| molecule_chembl_id | canonical_smiles | standard_value | binding_affinity |
|---|---|---|---|
| CHEMBL3678131 | Cc1c(C)n(Cc2cc(O[C@@H](C)C(=O)O)ccc2Cl)c2ccc(C(=O)N[C@@H](C)c3cccc(C(C)C)c3)cc12 | 0.06 | -10.208 |
| CHEMBL3695832 | CC[C@@H](NC(=O)c1ccc2c(c1)c(C)c(C)n2Cc1ccc(-c2ccccc2C(=O)O)cc1)c1ccccc1 | 0.13 | -11.55 |
| CHEMBL5191837 | Cc1cncc(-c2nc3cc(NC(=O)c4cc([N+](=O)[O-])ccc4Cl)ccc3o2)c1 | 0.16 | -8.04 |
| CHEMBL5207130 | CCc1ccc(-c2nc3cc(NC(=O)c4cc([N+](=O)[O-])ccc4Cl)ccc3o2)cc1 | 0.17 | -8.425 |
| CHEMBL3678134 | Cc1c(C)n(Cc2cc(O[C@@H](C)C(=O)O)ccc2Cl)c2ccc(C(=O)N[C@@H](C)c3ccc(C(C)(C)C)cc3)cc12 | 0.2 | -10.001 |
| CHEMBL3678128 | Cc1c(C)n(Cc2ccc(Cl)c(O[C@@H](C)C(=O)O)c2)c2ccc(C(=O)N[C@@H](C)c3ccc(C(C)(C)C)cc3)cc12 | 0.2 | -10.26 |
| CHEMBL5187164 | CCc1ccc(-c2nc3cc(NC(=O)c4cc([N+](=O)[O-])ccc4Br)ccc3o2)cc1 | 0.22 | -8.456 |
| CHEMBL5179281 | COc1ccc(-c2nc3cc(NC(=O)c4cc([N+](=O)[O-])ccc4Cl)ccc3o2)cc1F | 0.23 | -7.749 |

| CHEMBL5206512 | CCc1ccc(-c2nc3cc(NC(=O)c4cc([N+](=O)[O-])ccc4Cl)ccc3o2)cn1 | 0.25 | -8.428 |
| CHEMBL2088421 | COc1ccc(NC(=O)N(CCCCC2CCCCC2)CCc2ccc(SC(C)(C)C(=O)O)cc2)cc1 | 0.28 | -9.021 |

2D Descriptor Calculation

In this step, 2D molecular descriptors were calculated for each small molecule in the dataset, including the ligands and their binding affinities obtained from the docking simulations in Step 4. These 2D descriptors offer a numerical representation of the molecular structure, capturing various structural and physicochemical properties, making them ideal for input into machine learning models.

The calculation process began by deriving the 2D descriptors for each ligand. These descriptors were based on the SMILES notation of each molecule, which was processed to obtain values representing various molecular features, such as molecular weight, lipophilicity, and polarity. A variety of 2D descriptors were computed, covering topological, geometric, and physicochemical properties of the molecules.

After computing the 2D descriptors, a new column, titled *2d_descriptors*, was added to the dataset. This column contained the calculated descriptor values for each ligand. The updated dataset now included molecular identifiers, SMILES strings, IC50 values, docking binding affinities, and the newly computed 2D descriptors, all of which are crucial for subsequent machine learning analyses.

The final dataset, containing the 2D descriptors, was saved in the file *cleaned_ppar_docking_with_2d_descriptors.csv*. In this dataset, the *2d_descriptors* column represents a numerical vector of the calculated descriptors for each ligand, reflecting various molecular properties and structural features.

This dataset is now ready to be used as input for machine learning models, including deep learning and regression models, to predict the binding affinities of the ligands based on their structural and physicochemical characteristics.

Model Training and Evaluation

In this study, a deep learning model was developed to predict the binding affinity of ligands targeting the PPAR receptor family, based on their 2D descriptors. The model's performance was assessed using several key metrics, including Mean Squared Error (MSE), Mean Absolute Error (MAE), and R-squared (R²) values, for both the training and test datasets.

```
To focus on potent inhibitors, the dataset was filtered to include
only ligands with binding affinity values ≤ −5 kcal/mol. The
```

molecular structures of these ligands were represented by their 2D descriptors, derived from the SMILES notation of each molecule. These descriptors capture important molecular properties and structural features relevant to predicting binding affinity.

Before training, the binding affinity values were normalized using the StandardScaler to improve model convergence. The dataset was then divided into training and test sets, with 80% of the data allocated for training and 20% reserved for testing.

A deep learning model was built using the TensorFlow Keras library, designed to predict the binding affinity values of ligands as a regression task. The model architecture began with an input layer designed to accept 2D descriptor values, representing the ligands' structural features. This was followed by three fully connected hidden layers, containing 128, 64, and 64 neurons, respectively, with the ReLU activation function applied to introduce nonlinearity and enhance the model's learning capacity. The output layer consisted of a single neuron, responsible for predicting the binding affinity value.

The model was compiled using the Adam optimizer and the Mean Squared Error (MSE) loss function, which is suitable for regression tasks. Training was conducted over 50 epochs with a batch size of 32, and 20% of the dataset was reserved for validation to monitor the model's performance and prevent overfitting.

The model's performance evaluation showed its effectiveness in predicting binding affinities. The Mean Squared Error (MSE) was 0.361, while the Mean Absolute Error (MAE) was 0.436. The R-squared value for the training set was 0.862, indicating a strong fit, while the test set showed a value of 0.655, demonstrating good generalization. These results highlight the model's ability to capture the underlying patterns in the data.

For a visual assessment of the model's performance, a scatter plot comparing the true binding affinity values with the predicted binding affinities was generated. The plot included regression lines for both the training and test sets, further illustrating the model's ability to predict binding affinities accurately.

After training, the model was saved in the H5 format (*ppar_binding_affinity_model.h5*) for future use, including potential deployment in further studies or clinical applications.

A comparison plot of experimental binding affinity versus predicted binding affinity for the PPAR inhibitors was also generated and saved as *binding_affinity_comparison_plot.png* (Figure 2). The plot demonstrates a strong correlation between the true and predicted values, especially for the training set, emphasizing the model's effectiveness in predicting ligand binding affinity.

Figure 2. Experimental vs. Predicted Binding Affinity for PPAR Inhibitors.

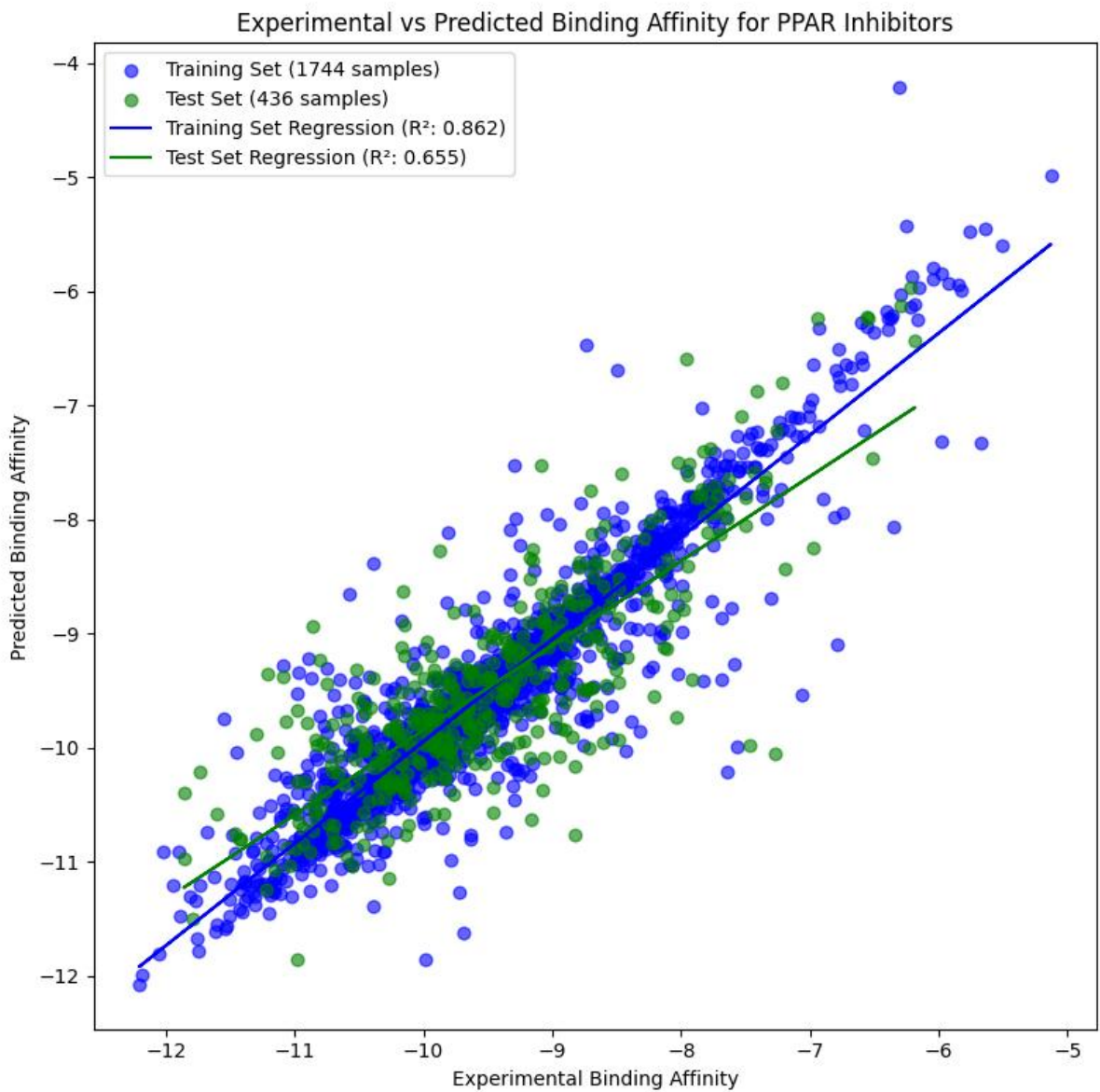

## Discussion

This study aimed to develop a deep learning model to predict the binding affinity of ligands targeting the PPAR receptor family, based on 2D molecular descriptors derived from SMILES notation. Ligand data were sourced from the ChEMBL database, providing a reliable and comprehensive foundation for the analysis. The dataset included compounds with known binding affinities, ensuring that the study concentrated on experimentally validated inhibitors. This selection allowed for confident downstream analyses, ensuring the data were relevant for predicting ligand-receptor interactions.

## Strengths of the Dataset

The ChEMBL database is a widely trusted resource, known for its comprehensive and high-quality bioactivity data, making it an excellent source for ligand information. The compounds chosen for this study had strong experimental validation, providing confidence in their interactions with the PPAR receptor. This feature of the dataset enhanced the credibility of our predictions, ensuring that the results were grounded in reliable bioactivity data.

**Challenges and Data Cleaning**

While the ChEMBL dataset provided a solid foundation, significant preprocessing was required due to missing or invalid data for some ligands, such as missing binding affinities. These discrepancies in the dataset necessitated extensive cleaning. Duplicate entries were removed, and invalid or incomplete data points were excluded. Following this process, the dataset was reduced from 15,161 to 9,947 ligands with valid IC50 values, ensuring that the data maintained its integrity and suitability for model training.

Though this cleaning step reduced the sample size, it was essential to prioritize the inclusion of only high-quality, reliable data to avoid introducing bias or inaccuracies into the model. Despite the reduction, this approach enhanced the quality of the training dataset, ensuring more accurate and trustworthy results.

**2D Descriptors as Molecular Representations**

In this study, 2D molecular descriptors were employed to represent the ligands' molecular structures. Unlike more complex representations such as 3D molecular structures or fingerprints, 2D descriptors offer a simplified yet effective means to capture key molecular features such as atom types, bond types, and functional groups. These descriptors are widely used in cheminformatics, facilitating molecular similarity searches and feature extraction for machine learning models.

The conversion of SMILES representations into 2D descriptors allowed the generation of a compact and informative set of features for model training. Although 2D descriptors are efficient in capturing structural information, they may not account for all relevant chemical features, particularly those involved in more complex molecular interactions. The choice of descriptor parameters, such as the number and type of descriptors, plays a significant role in the model's ability to predict binding affinities accurately.

**Molecular Docking to Predict Binding Affinities**

To predict the binding affinities of the ligands for the PPAR receptor, molecular docking simulations using AutoDock Vina were employed. This tool is widely used in computational chemistry for its efficiency and accuracy in predicting ligand-receptor binding interactions. The docking results provided predicted binding affinities, which were integral for training the predictive model.

Although molecular docking is a valuable method for affinity prediction, it is subject to certain limitations. The accuracy of docking results can be influenced by factors such as receptor structure quality, docking parameters, and the precision of the binding site used in the simulations. While the receptor was adequately prepared for docking in this study, real-world applications may face challenges in maintaining docking accuracy due to these factors.

**Model Development and Evaluation**

The central task of this study involved developing a deep learning model to predict the binding affinity of PPAR inhibitors. The model was trained using 2D descriptors as input features and normalized binding affinities as output. The deep learning model consisted of three hidden layers, each using the ReLU activation function to capture complex relationships between the molecular features and predicted binding affinity. Model performance was evaluated using key metrics including R-squared ($R^2$), Mean Squared Error (MSE), and Mean Absolute Error (MAE), all of which provide insight into the accuracy and reliability of the model's predictions.

**Performance Analysis**

The model demonstrated strong performance on the training dataset, with an $R^2$ value of 0.862, suggesting that it explained most of the variance in the binding affinity data. However, the $R^2$ value on the test set dropped to 0.655, indicating that the model was able to generalize reasonably well to new, unseen data. The drop in performance from training to test data is common in machine learning models, especially when the model is complex or trained on a limited dataset, and suggests there is potential for improvement, particularly in enhancing generalization.

The Mean Squared Error (MSE) value of 0.361 indicates that the model's predictions were relatively close to the actual binding affinities on average. However, MSE is sensitive to outliers, meaning that a few extreme errors may have contributed to the value. To further improve the model, addressing these outliers or refining the model's handling of complex molecular interactions might be necessary.

The Mean Absolute Error (MAE) value of 0.436 further supports the model's reliability, as it shows that the average prediction error is relatively small. Since MAE is less sensitive to outliers than MSE, it serves as a robust indicator of overall prediction accuracy. Although the MAE suggests that the model is generally reliable, there is room for improvement to reduce prediction errors further.

The high $R^2$ value for the training set (0.862) indicates that the 2D descriptors used as features were highly informative in learning the relationships between the ligands' molecular structures and their binding affinity. The $R^2$ value of 0.655 on the test set shows that the model generalizes well to new data, although the drop between the training and test sets suggests that model refinement could improve its ability to predict binding affinity more consistently across diverse datasets.

**Conclusion**

This study successfully developed a deep learning model to predict the binding affinity of ligands targeting the PPAR receptors, using 2D molecular descriptors derived from SMILES notation. The model was trained using data from the ChEMBL database, which provides information on ligands with experimentally validated binding affinities. Through rigorous data cleaning, the dataset used for training contained 9,947 ligands with valid IC50 values, ensuring the quality of the data used for model training.

2D molecular descriptors were employed to represent the molecular structure of the ligands, providing an effective representation of their structural and physicochemical features. Molecular docking results using AutoDock Vina provided the binding affinities used to train the predictive model. The developed deep learning model demonstrated good performance in predicting binding affinities, with an $R^2$ value of 0.862 on the training set and 0.655 on the test set. Although there was a decrease in performance between the training and test sets, the model still showed good generalization capability on previously unseen data.

Evaluation results using MSE and MAE indicate that the model can predict binding affinities with reasonable accuracy, though some predictions exhibited larger errors that need further refinement. Overall, the model shows strong potential in predicting PPAR ligand-receptor interactions and can be applied to further research for more effective ligand-based drug design.

Moving forward, model improvements, including the selection of molecular descriptors and handling of extreme data, can enhance prediction accuracy and expand the model's capabilities for real-world applications, such as drug development for diseases related to PPAR.

**Acknowledgments**

We would like to express our sincere gratitude to the Fugaku Supercomputer facility for providing the essential computational resources that enabled the molecular docking simulations conducted in this study. The high-performance computing capabilities of Fugaku were crucial in efficiently calculating the binding affinities for ligands targeting the PPAR receptor family. This research would not have been possible without their generous support. We also extend our thanks to the ChEMBL database for supplying the valuable data that formed the cornerstone of this work.

**Conflicts of Interest Statement**

The authors declare no conflicts of interest.

**Authors' Contributions**

All authors equally contributed to the conception, design, and execution of this study. All authors have read and approved the final version of the manuscript.

Appendix 1

```python
import os
import pandas as pd
from chembl_webresource_client.new_client import new_client

def search_targets(query_list, job_dir):
    for search_term in query_list:
        # Load target data
        target = new_client.target

        # Perform target search
        try:
            target_query = target.search(search_term)

            # Pastikan ada data yang ditemukan
            if target_query:
                # Convert to DataFrame
                targets = pd.DataFrame.from_dict(target_query)

                # Pilih kolom yang relevan dan tambahkan kolom search_term
                search_term_formatted = search_term.replace(" ", "-")
                selected_columns = ['target_chembl_id', 'pref_name']
                targets['search_term'] = search_term_formatted
                targets = targets[selected_columns]

                # Pastikan direktori tujuan ada
                if not os.path.exists(job_dir):
                    os.makedirs(job_dir)

                # Save DataFrame ke CSV
                targets_output_file_path = os.path.join(job_dir, f'targets_{search_term_formatted}.csv')
                targets.to_csv(targets_output_file_path, index=False)
                print(f"Results for '{search_term}' saved to {targets_output_file_path}")
            else:
                print(f"No results found for '{search_term}'")

        except Exception as e:
            print(f"Error occurred while searching for '{search_term}': {e}")

def main():
    # Definisikan query_list langsung di dalam kode
    query_list = ['alpha_glucosidase']  # Gantilah ini dengan kata kunci yang diinginkan
```

```python
    job_dir = os.getcwd()  # Menggunakan direktori kerja saat ini

    # Menjalankan pencarian target
    search_targets(query_list, job_dir)

if __name__ == "__main__":
    main()
```

Appendix 2

```python
import pandas as pd
import numpy as np
import os
from chembl_webresource_client.new_client import new_client
from rdkit import Chem
from tqdm import tqdm
from concurrent.futures import ProcessPoolExecutor
from glob import glob

# Function to validate smiles
def validate_smiles(smiles):
    mol = Chem.MolFromSmiles(smiles)
    if mol is not None:
        return Chem.MolToSmiles(mol)
    else:
        return None

def combine_csv_files(target, target_dir):
    # Read all CSV files in the directory
    csv_files = glob(os.path.join(target_dir, '*.csv'))

    if not csv_files:
        print("No CSV files found in the directory.")
        return None  # Return None if no files found

    # Read each file and combine them into one DataFrame
    dfs = [pd.read_csv(file) for file in csv_files if os.path.getsize(file) > 0]

    if not dfs:
        print(f"No data in CSV files in the directory {target_dir}.")
        return None  # Return None if no data in files

    combined_df = pd.concat(dfs, ignore_index=True)
    output_file = os.path.join(target_dir, f'{target}.csv')

    # Save the combined DataFrame to one CSV file
    combined_df.to_csv(output_file, index=False)
    print(f"Data {target_dir} saved in {output_file}")
    return combined_df

def process_chembl_id(chembl_id, target, target_dir, other, others_column):
    try:
        # Load data
```

```python
        activity = new_client.activity
        res = activity.filter(target_chembl_id=chembl_id).filter(standard_type="IC50")
        df = pd.DataFrame.from_dict(res)

        # Extract columns
        if 'molecule_chembl_id' in df.columns:
            mol_cid = list(df.molecule_chembl_id)
        else:
            mol_cid = []

        if 'canonical_smiles' in df.columns:
            canonical_smiles = list(df.canonical_smiles)
        else:
            canonical_smiles = []

        if 'standard_value' in df.columns:
            standard_value = list(df.standard_value)
        else:
            standard_value = []

        # Create DataFrame with all columns
        data_tuples = list(zip(mol_cid, canonical_smiles, standard_value))
        df = pd.DataFrame(data_tuples, columns=['molecule_chembl_id', 'canonical_smiles', 'standard_value'])

        # Clean and remove duplicates
        df = df.dropna(subset=['molecule_chembl_id'])
        df = df.dropna(subset=['canonical_smiles'])
        df['canonical_smiles'] = df['canonical_smiles'].apply(validate_smiles)
        df = df.drop_duplicates(subset=['molecule_chembl_id'], keep='first', ignore_index=True)

        # Add a new column others and 'target' with the value {target}
        df[others_column] = other
        df['search_term'] = target

        output_file_path = os.path.join(target_dir, f'{chembl_id}.csv')
        df.to_csv(output_file_path, index=False)

        return f"Processed {chembl_id}"

    except Exception as e:
        return f"Error processing {chembl_id}: {str(e)}"
```

```python
def search_molecules(job_dir):
    # Set the random seed for reproducibility
    np.random.seed(1)

    df_all_target = []
    targets = []
    all_results = []

    targets_files = glob(os.path.join(job_dir, 'targets_*.csv'))

    for targets_file_path in targets_files:
        df_targets = pd.read_csv(targets_file_path)
        target_chembl_ids = df_targets['target_chembl_id']
        others_columns = df_targets.columns.difference(['target_chembl_id'])  # Get columns other than 'target_chembl_id'
        target = os.path.splitext(os.path.basename(targets_file_path))[0].replace("targets_", "")
        target_dir = os.path.join(job_dir, target)
        targets.append(target)

        if not os.path.exists(target_dir):
            os.mkdir(target_dir)

        max_workers = os.cpu_count()
        futures = []

        with ProcessPoolExecutor(max_workers=max_workers) as executor:
            # Process chembl_ids in parallel
            for chembl_id in target_chembl_ids:
                other_values = df_targets[df_targets['target_chembl_id'] == chembl_id][others_columns].iloc[0]

                future = executor.submit(process_chembl_id, chembl_id, target, target_dir, other_values, others_columns)
                futures.append(future)

            # Wait for all tasks to complete
            results = [future.result() for future in futures]

        combined_df = combine_csv_files(target, target_dir)
        print(combined_df)

        if not combined_df.empty:  # Check if DataFrame is not empty before appending
            df_all_target.append(combined_df)
            # Remove the following lines as DataFrame doesn't have 'head' and 'info' methods
```

```python
            print(df_all_target)

    if df_all_target:
        # Combine all DataFrames in the list
        df_all_target_combined = pd.concat(df_all_target, ignore_index=True)

        # Save the combined DataFrame to one CSV file
        csv_molecules = os.path.join(job_dir, 'merging_' + '_'.join(targets) + '.csv')
        df_all_target_combined.to_csv(csv_molecules, index=False)

    return csv_molecules

if __name__ == "__main__":
    job_dir = os.getcwd()
    search_molecules(job_dir)
```

Appendix 3

```python
import pandas as pd

def clean_combined_data(input_file, output_file):
    """
    Cleans the combined CSV data by:
    1. Removing duplicate rows based on 'molecule_chembl_id',
       keeping the row with the lowest 'standard_value'.
    2. Dropping rows where 'standard_value' is NaN or 0.
    3. Dropping the columns 'pref_name' and 'search_term'.

    Args:
        input_file (str): Path to the input CSV file.
        output_file (str): Path to save the cleaned CSV file.

    Returns:
        str: Path to the cleaned file.
    """
    # Load the CSV file into a DataFrame
    df = pd.read_csv(input_file)

    # Drop rows where 'standard_value' is NaN or 0
    if 'standard_value' in df.columns:
        df['standard_value'] = pd.to_numeric(df['standard_value'], errors='coerce')  # Ensure standard_value is numeric
        df = df[df['standard_value'].notna() & (df['standard_value'] != 0)]  # Remove rows where standard_value is NaN or 0

    # Drop duplicates by molecule_chembl_id, keeping the row with the lowest standard_value
    df = df.sort_values('standard_value').drop_duplicates(subset='molecule_chembl_id', keep='first')

    # Drop the columns 'pref_name' and 'search_term' if they exist
    columns_to_drop = ['pref_name', 'search_term']
    df = df.drop(columns=[col for col in columns_to_drop if col in df.columns])

    # Save the cleaned DataFrame to a new CSV file
    df.to_csv(output_file, index=False)
    print(f"Cleaned data saved to: {output_file}")
    return output_file

# Example usage
if __name__ == "__main__":
    input_file = "ppar.csv"  # Path to the input file
```

```python
output_file = "ppar_cleaned.csv"  # Path to save the cleaned file

clean_combined_data(input_file, output_file)
```

```python
import os
import pandas as pd
from rdkit import Chem
from rdkit.Chem import AllChem
from subprocess import run, DEVNULL, CalledProcessError
from tqdm import tqdm
from concurrent.futures import ProcessPoolExecutor

def smiles_to_pdbqt(smiles, ligand_name):
    try:
        mol = Chem.MolFromSmiles(smiles)
        if mol is None:
            raise ValueError(f"Invalid SMILES: {smiles}")

        mol = Chem.AddHs(mol)
        AllChem.EmbedMolecule(mol, AllChem.ETKDG())
        AllChem.UFFOptimizeMolecule(mol)

        pdb_path = f"{ligand_name}.pdb"
        Chem.MolToPDBFile(mol, pdb_path)

        pdbqt_path = f"{ligand_name}.pdbqt"
        run(["obabel", pdb_path, "-O", pdbqt_path, "--gen3d"], stdout=DEVNULL, stderr=DEVNULL, check=True)

        os.remove(pdb_path)  # Hapus file PDB setelah konversi
        return pdbqt_path
    except (ValueError, CalledProcessError, Exception) as e:
        print(f"Error converting {ligand_name} to PDBQT: {e}")
        return None

def run_vina_docking(receptor, config_file, ligand_pdbqt, ligand_name):
    try:
        output_pdbqt = f"{ligand_name}_out.pdbqt"

        run([
            "vina",
            "--receptor", receptor,
            "--ligand", ligand_pdbqt,
            "--config", config_file,
            "--out", output_pdbqt
```

```python
        ], stdout=DEVNULL, stderr=DEVNULL, check=True)

        return output_pdbqt
    except CalledProcessError as e:
        print(f"Vina docking failed for {ligand_name}: {e}")
        return None

def extract_best_binding_affinity(output_pdbqt):
    try:
        with open(output_pdbqt, 'r') as file:
            for line in file:
                if "REMARK VINA RESULT" in line:
                    return float(line.split()[3])  # Mengambil nilai affinity terbaik
    except Exception as e:
        print(f"Error reading output file {output_pdbqt}: {e}")
    return None

def process_ligand(row, receptor, config_file):
    smiles = row['canonical_smiles']
    molecule_id = row['molecule_chembl_id']
    standard_value = row['standard_value']
    ligand_name = f"ligand_{molecule_id}"

    ligand_pdbqt = smiles_to_pdbqt(smiles, ligand_name)
    if not ligand_pdbqt:
        return molecule_id, smiles, standard_value, None

    output_pdbqt = run_vina_docking(receptor, config_file, ligand_pdbqt, ligand_name)
    best_affinity = None
    if output_pdbqt:
        best_affinity = extract_best_binding_affinity(output_pdbqt)
        os.remove(output_pdbqt)

    if os.path.exists(ligand_pdbqt):
        os.remove(ligand_pdbqt)

    return molecule_id, smiles, standard_value, best_affinity

def perform_docking(input_file, receptor, config_file, output_csv):
    if not os.path.exists(receptor):
        print(f"Receptor file not found: {receptor}")
```

```python
        return

    if not os.path.exists(config_file):
        print(f"Config file not found: {config_file}")
        return

    input_df = pd.read_csv(input_file)
    if 'canonical_smiles' not in input_df.columns or 'molecule_chembl_id' not in input_df.columns or 'standard_value' not in input_df.columns:
        print("Required columns 'canonical_smiles', 'molecule_chembl_id', or 'standard_value' missing in input file.")
        return

    # Pastikan file output CSV memiliki header jika belum ada
    if not os.path.exists(output_csv):
        with open(output_csv, 'w') as f:
            f.write("molecule_chembl_id,canonical_smiles,standard_value,binding_affinity\n")

    # Proses secara paralel
    rows = input_df.to_dict('records')
    with ProcessPoolExecutor(max_workers=1) as executor:
        futures = [executor.submit(process_ligand, row, receptor, config_file) for row in rows]

        for future in tqdm(futures, total=len(rows), desc="Docking ligands"):
            try:
                result = future.result()
                if result:
                    molecule_id, smiles, standard_value, best_affinity = result
                    # Tambahkan hasil ke file CSV output
                    with open(output_csv, 'a') as f:
                        f.write(f"{molecule_id},{smiles},{standard_value},{best_affinity}\n")
            except Exception as e:
                print(f"Error during docking: {e}")

    print(f"Docking results saved to {output_csv}")

if __name__ == "__main__":
    input_file = "ppar_cleaned.csv"
    receptor = "ppar.pdbqt"
    config_file = "config.txt"
    output_csv = "ppar_docking.csv"

    perform_docking(input_file, receptor, config_file, output_csv)
```

Appendix 5

```python
import pandas as pd
from rdkit import Chem
from rdkit.Chem import Descriptors

# Load the dataset
df = pd.read_csv('ppar_docking.csv')

# Function to generate 2D descriptors from SMILES using the provided get_2d_descriptors function
def get_2d_descriptors(mol, missing_val=None):
    descriptors_2d = {}
    if mol is not None:
        try:
            # Iterate over all the descriptors in RDKit's Descriptors.descList
            for desc_name, fn in Descriptors.descList:
                val = fn(mol)
                descriptors_2d[desc_name] = val

        except Exception as e:
            print(f"Error calculating 2D descriptors for: {mol}")
            descriptors_2d = {desc_name: missing_val for desc_name, _ in Descriptors.descList}
    else:
        descriptors_2d = {desc_name: missing_val for desc_name, _ in Descriptors.descList}
        print(f"Molecule is None for: {mol}")

    # Return the descriptors as a DataFrame
    descriptors_df = pd.DataFrame([descriptors_2d])
    return descriptors_df

# Function to convert SMILES to 2D descriptors
def generate_2d_descriptors(smiles):
    mol = Chem.MolFromSmiles(smiles)
    if mol:
        # Use the get_2d_descriptors function to calculate descriptors
        return get_2d_descriptors(mol)
    else:
        return None

# Apply the function to each molecule in the dataset
df_descriptors = df['canonical_smiles'].apply(generate_2d_descriptors)

# Concatenate the descriptors DataFrame with the original dataset
# Convert list of descriptors DataFrame into a single DataFrame
```

```python
descriptors_df = pd.concat([x for x in df_descriptors if x is not None], ignore_index=True)

# Combine the original dataset with the descriptors
df = pd.concat([df, descriptors_df], axis=1)

# Save the dataset with the new 2D descriptors
df.to_csv('ppar_2d_descriptors.csv', index=False)

# Display first few rows of the dataframe
print(df.head())
```

```python
import pandas as pd
import numpy as np
from sklearn.model_selection import train_test_split
from sklearn.preprocessing import StandardScaler
from sklearn.metrics import mean_squared_error, r2_score
import tensorflow as tf
from tensorflow.keras import layers, models
import matplotlib.pyplot as plt
from sklearn.linear_model import LinearRegression

# Load the dataset
df = pd.read_csv('ppar_2d_descriptors.csv')

# Filter rows where binding_affinity <= -5
df_filtered = df[df['binding_affinity'] <= -5]

# Drop non-feature columns ('molecule_chembl_id', 'canonical_smiles', 'standard_value', 'binding_affinity')
X = df_filtered.drop(columns=['molecule_chembl_id', 'canonical_smiles', 'standard_value', 'binding_affinity'])

# Handle NaN values in the features (X)
# Option 1: Drop columns with NaN values (remove columns with missing data)
X = X.dropna(axis=1)

# Option 2: Fill NaN values with the mean (or median) of each column
# X = X.fillna(X.mean())  # Fill NaN with the mean of each column

# Extract the target variable (binding_affinity)
y = df_filtered['binding_affinity'].values

# Check if y has NaN values and handle them (if necessary)
y = np.nan_to_num(y)  # Replace NaN with 0 or you can use another strategy like the mean

# Ensure that X and y have the same length
assert len(X) == len(y), f"Length of X: {len(X)} and y: {len(y)} do not match."

# Normalize the features (X) and target (y) separately
scaler_X = StandardScaler()
X_scaled = scaler_X.fit_transform(X)  # Normalize the features

scaler_y = StandardScaler()
```

```python
y_scaled = scaler_y.fit_transform(y.reshape(-1, 1)).flatten()  # Normalize the target (binding affinity)

# Split the dataset into training and testing sets
X_train, X_test, y_train, y_test = train_test_split(X_scaled, y_scaled, test_size=0.2, random_state=42)

# Print the number of samples in training and test sets
num_train_samples = len(X_train)
num_test_samples = len(X_test)
print(f'Number of molecules in Training Set: {num_train_samples}')
print(f'Number of molecules in Test Set: {num_test_samples}')

# Build the deep learning model
model = models.Sequential()
model.add(layers.InputLayer(input_shape=(X_train.shape[1],)))  # Input layer
model.add(layers.Dense(128, activation='relu'))  # First hidden layer
model.add(layers.Dense(64, activation='relu'))   # Second hidden layer
model.add(layers.Dense(64, activation='relu'))   # Third hidden layer
model.add(layers.Dense(1))  # Output layer (regression)

# Compile the model
model.compile(optimizer='adam', loss='mean_squared_error', metrics=['mae'])

# Train the model
history = model.fit(X_train, y_train, epochs=50, batch_size=32, validation_split=0.2, verbose=1)

# Predict on training and test data
y_pred_train_scaled = model.predict(X_train)
y_pred_test_scaled = model.predict(X_test)

# Inverse transform the predicted values to the original scale (for y)
y_pred_train = scaler_y.inverse_transform(y_pred_train_scaled.reshape(-1, 1)).flatten()
y_pred_test = scaler_y.inverse_transform(y_pred_test_scaled.reshape(-1, 1)).flatten()

# Inverse transform the true values to the original scale (for y)
y_train_original = scaler_y.inverse_transform(y_train.reshape(-1, 1)).flatten()
y_test_original = scaler_y.inverse_transform(y_test.reshape(-1, 1)).flatten()

# Inverse transform the features (X) for visualization (if required)
X_train_original = scaler_X.inverse_transform(X_train)
X_test_original = scaler_X.inverse_transform(X_test)
```

```python
# Calculate R-squared for both train and test sets
r2_train = r2_score(y_train_original, y_pred_train)
r2_test = r2_score(y_test_original, y_pred_test)

# Calculate evaluation metrics
mse = mean_squared_error(y_test_original, y_pred_test)
mae = np.mean(np.abs(y_test_original - y_pred_test))

print(f"Mean Squared Error: {mse}")
print(f"Mean Absolute Error: {mae}")
print(f"R-squared (Training Set): {r2_train}")
print(f"R-squared (Test Set): {r2_test}")

# Optionally, save the trained model
model.save('ppar_2d_descriptors_model.keras')

# Create LinearRegression models for the training and test sets to plot regression lines
train_lr = LinearRegression()
test_lr = LinearRegression()

# Train linear models
train_lr.fit(y_train_original.reshape(-1, 1), y_pred_train)
test_lr.fit(y_test_original.reshape(-1, 1), y_pred_test)

# Predict using the linear models
train_line = train_lr.predict(y_train_original.reshape(-1, 1))
test_line = test_lr.predict(y_test_original.reshape(-1, 1))

# Plotting the comparison between true and predicted values (combined for train and test)
plt.figure(figsize=(8, 8))

# Scatter plot for training set and test set
plt.scatter(y_train_original, y_pred_train, color='blue', alpha=0.6, label=f'Training Set ({num_train_samples} samples)', s=40)
plt.scatter(y_test_original, y_pred_test, color='green', alpha=0.6, label=f'Test Set ({num_test_samples} samples)', s=40)

# Add the regression line for training set
plt.plot(y_train_original, train_line, color='blue', linestyle='-', label=f'Training Set Regression (R²: {r2_train:.3f})')

# Add the regression line for test set
plt.plot(y_test_original, test_line, color='green', linestyle='-', label=f'Test Set Regression (R²: {r2_test:.3f})')
```

```python
# Add labels and title with R-squared values in the title
plt.title(f'Experimental vs Predicted Binding Affinity for PPAR Inhibitors')
plt.xlabel('Experimental Binding Affinity')
plt.ylabel('Predicted Binding Affinity')

# Show legend
plt.legend()

# Save the plot to a file (PNG format)
plt.tight_layout()
plt.savefig('ppar_2d_descriptors_plot.png')  # Save as PNG

# Show the plot
plt.show()
```